\documentclass[preprint,showpacs,preprintnumbers,nofootinbib,dvipdfmx]{revtex4-1}

\usepackage{graphicx}
\usepackage{color}

\newcommand{\lsim}{\raise0.3ex\hbox{$\;<$\kern-0.75em\raise-1.1ex\hbox{$\sim\;$}}}
\newcommand{\gsim}{\raise0.3ex\hbox{$\;>$\kern-0.75em\raise-1.1ex\hbox{$\sim\;$}}}

\newcommand{\lmlt}{L_\mu^{} - L_\tau^{}}
\newcommand{\Zp}{Z_{\mu\tau}^{}}
\newcommand{\bmx}{\left(\begin{array}}
\newcommand{\emx}{\end{array}\right)}

\def\vev#1{\langle{#1}\rangle}

\begin{document}

\title{
Low scale seesaw models for low scale $U(1)_{\lmlt}^{}$ symmetry
}

\author{Takeshi Araki}
\email{kt13579@ns.kogakuin.ac.jp}
\affiliation{%
Learning Support Center, Kogakuin University,
\\
2665-1 Nakano, 
Hachioji, 
192-0015 Tokyo, 
Japan
}

\author{Kento Asai}
\email{asai@hep-th.phys.s.u-tokyo.ac.jp}
\affiliation{%
Department of Physics, University of Tokyo,
\\
Bunkyo-ku, Tokyo 133-0033, Japan
}

\author{Joe Sato}
\email{joe@phy.saitama-u.ac.jp}
\affiliation{%
Department of Physics, Saitama University,
\\
Shimo-Okubo 255, 
338-8570 Saitama Sakura-ku, 
Japan
}

\author{Takashi Shimomura}
\email{shimomura@cc.miyazaki-u.ac.jp}
\affiliation{%
Faculty of Education, 
University of Miyazaki,
\\
1-1 Gakuen-Kibanadai-Nishi,
889-2192 Miyazaki,
Japan
}

\date{\today}


\preprint{\bf STUPP-19-237, UME-PP-010, UT-19-22}


\begin{abstract}
We propose models for neutrino masses and mixing in the framework of low scale $U(1)_{\lmlt}^{}$ gauge extension of the standard model.
The models are designed to spontaneously break $U(1)_{\lmlt}^{}$ so that the $U(1)_{\lmlt}^{}$ gauge boson acquires an MeV scale mass, which is required to solve the long-standing problem of muon anomalous magnetic moment.
Tiny neutrino masses are obtained by simultaneously invoking the linear and the inverse seesaw mechanism, and we succeed in realizing two types of one-zero textures in the active neutrino mass matrix.
Both of the obtained textures favor inverted neutrino mass ordering and are testable in next generation experiments of neutrinoless double beta decay.
We also show that some of extra scalar bosons can have MeV scale masses and would have significant impacts on observations of high energy cosmic neutrinos.
\end{abstract}

\maketitle

\section{Introduction}
The extension of the standard model (SM) is one of the highest priority issues in modern particle physics; various types of new physics models have been proposed in the literature.
In most cases, new physics is anticipated to exist at an energy scale much higher than the electroweak scale, because it is strictly constrained by low energy experiments.
Nevertheless, it may also be possible to consider new physics at lower energy, if its interactions are sufficiently weak so that the experimental constraints can be evaded.
One interesting possibility in that direction is gauging the muon number minus the tau number, that is, the so-called $U(1)_{\lmlt}$ gauge extension \cite{Foot:1990mn,He:1990pn,Foot:1994vd}.
This could be one of the most natural extensions of the SM since it is gauge anomaly free within the SM particle contents.
Furthermore, it was recently found, in Ref. \cite{Altmannshofer:2014pba} (see also Refs. \cite{Gninenko:2001hx,Baek:2001kca}), that the $U(1)_{\lmlt}^{}$ gauge boson, $Z_{\mu\tau}^{}$, with an MeV scale mass can settle the long-standing discrepancy of muon anomalous magnetic moment ($g_\mu - 2$) \cite{Bennett:2006fi,Melnikov:2003xd,Davier:2010nc,Hagiwara:2011af,Aoyama:2012wk,Kurz:2014wya} without conflicting other experimental constraints.
Then, this result motivated many authors to study low scale $U(1)_{\lmlt}$ extension of the SM in various contexts: for instance, 
it was found that an MeV scale $\Zp$ can relax the tension between the late time and the early time determination of the Hubble constant \cite{Escudero:2019gzq}, 
implications for the dark matter problem were studied in Refs. \cite{Baek:2015fea,Biswas:2016yjr,Kamada:2018zxi,Foldenauer:2018zrz}, 
and the detectability of such a light $\Zp$ was discussed in Refs. \cite{Gninenko:2014pea,Kaneta:2016uyt,Araki:2017wyg,Chen:2017cic,Gninenko:2018tlp,Chun:2018ibr,Nomura:2019uyz,Jho:2019cxq}.
Moreover, it was pointed out in Refs. \cite{Araki:2014ona,Kamada:2015era,DiFranzo:2015qea,Araki:2015mya} that an MeV scale $\Zp$ causes significant attenuation in the flux of high energy cosmic neutrinos and can explain the unexpected dip in the energy spectrum of high energy cosmic neutrinos reported by the IceCube Collaboration.

From a viewpoint of the lepton mixing, the $U(1)_{\lmlt}$ symmetry is well known to naturally explain the observed large atmospheric mixing angle \cite{Ma:2001md,Choubey:2004hn,Ota:2006xr}.
However, it is also well known that the exact $U(1)_{\lmlt}$ symmetry forbids many of entries in the neutrino mass matrix, and as a result the solar and the reactor mixing angle are forced to be zero. 
In order to remedy such a situation, extra scalars are often introduced to spontaneously break $U(1)_{\lmlt}$ and to revive some of the entries forbidden by $U(1)_{\lmlt}^{}$, which provides us with an opportunity to realize zero textures in the neutrino mass matrix and testability of the model.
Especially, the type-C \cite{Frampton:2002yf} two-zero texture or two-zero-minor structure is frequently obtained \cite{Heeck:2011wj,Dev:2017fdz,Nomura:2018vfz,Araki:2012ip,Biswas:2016yan,Asai:2017ryy,Chiang:2017vcl,Nomura:2018cle,Asai:2018ocx}, 
and these structures were consistent with experiments at that time.
However, in Ref. \cite{Asai:2018ocx}, it has been pointed out that the type-C two-zero-minor structure is now driven into a corner by the combined upper bound on the sum of neutrino masses placed by the Planck Collaboration: $\sum_i m_i <0.12$ eV \cite{Aghanim:2018eyx}.
Similarly, it can be checked that the type-C two-zero texture is ruled out by the Planck bound\footnote{
See Fig. 9 in Ref. \cite{Dev:2017fdz} and Fig. 4 in Ref. \cite{Singh:2019baq}.
We have done the same calculation and obtained a similar conclusion that the sum of neutrino masses is constrained to be larger than $0.14$ eV within $3\sigma$ error ranges of the neutrino oscillation parameters.
}.

In this work, we improve the previous studies mentioned above and propose experimentally consistent models for $U(1)_{\lmlt}^{}$ extension of the SM.
For this purpose, we combine the inverse and the linear seesaw mechanism and succeed in realizing two types of one-zero textures 
\cite{Xing:2003ic,Merle:2006du,Lashin:2011dn,Deepthi:2011sk,Gautam:2015kya,Kitabayashi:2018bye}.
Both of the obtained textures prefer inverted neutrino mass ordering and are consistent with the Planck bound as well as the bounds from neutrino oscillation experiments.
We calculate the effective mass of neutrinoless double beta decay as our prediction and find that it is testable in next generation experiments.
In the models, extra scalars are introduced to spontaneously break $U(1)_{\lmlt}^{}$ and to give an MeV scale mass to $\Zp$ in order that the $g_\mu-2$ problem and the dip in the IceCube data can simultaneously be solved.
We show that some of the extra scalar bosons can have MeV scale masses and can attenuate the flux of high energy cosmic neutrinos, just as the $\Zp$ does.

The paper is organized as follows.
In Sec. \ref{sec:model}, we describe the particle contents and the charge assignments of the models.
Then, in Sec. \ref{sec:neutrino}, we show that tiny neutrino masses are obtained by combining the linear and the inverse seesaw mechanism.
The numerical calculations of neutrino masses and mixing are given in Sec. \ref{sec:num}.
The scalar sector is studied in Sec. \ref{sec:scalar}, in which we also show that some scalar bosons can have MeV scale masses.
In Sec. \ref{sec:SnI}, we briefly discuss whether the MeV scale scalar bosons can attenuate the flux of high energy cosmic neutrinos.
Finally, we summarize our results in Sec. \ref{sec:sum}.

\section{Models}
\label{sec:model}
\begin{table}[h]
\begin{tabular}{|c|c|c|c|c|c|c|c|c|}\hline
  & 
$\ell_{L_e},\ell_{L_\mu},\ell_{L_\tau}$ & 
$\ell_{R_e},\ell_{R_\mu},\ell_{R_\tau}$ & 
$N_{R_e},N_{R_{\mu(\tau)}}$ & 
$N_{L_e},N_{L_{\mu(\tau)}}$ & 
$H$ & $\Phi$ & $S_L$ & $S_{\mu\tau}$ \\ \hline
$U(1)_{L_\mu - L_\tau}^{}$ & $0,1,-1$ & $0,1,-1$ & $0,1(-1)$ & $0,1(-1)$ & $0$ & $0$ & $0$ & $1$ \\ \hline
$U(1)_L$ & $1$ & $1$ & $1$ & $1$ & $0$ & $-2$ & $-2$ & $0$ \\ \hline
\end{tabular}
\caption{The charge assignments of $U(1)_{L_\mu - L_\tau}^{}$ and $U(1)_L^{}$ for Model-A(B).}
\label{tab1}
\end{table}
We begin with $U(1)_{\lmlt}^{}$ extension of the SM and introduce left- and right-handed SM gauge singlet fermions, $N_L$ and $N_R$.
Although it may be natural to introduce three generations of $N_R$ and $N_L$ and assign them $U(1)_{L_\mu-L_\tau}^{}$ charges as $(N_{R(L)_e},~N_{R(L)_\mu},~N_{R(L)_\tau})=(0,1,-1)$.
Nevertheless, in this work, we introduce only two generations and aim at building a minimal model.
In this case, the following two possibilities can be considered, and we refer to them as Model-A and Model-B:
\begin{eqnarray*}
&&
{\rm Model-A} ~~\ddag~~ (N_{R(L)_e},~N_{R(L)_\mu})=(0,1),
\nonumber \\
&&
{\rm Model-B} ~~\ddag~~ (N_{R(L)_e},~N_{R(L)_\tau})=(0,-1)
.
\end{eqnarray*}
Note that we omit the case of $(N_{R(L)_\mu},~N_{R(L)_\tau})=(1,-1)$ because it results in a neutrino mass matrix of
\begin{eqnarray}
m_\nu =
\bmx{ccc}
0 & 0 & 0 \\
0 & 0 & m \\
0 & m & 0
\emx ,
\end{eqnarray}
which predicts unrealistic neutrino mass spectrum and mixing.
Note also that the other combinations give rise to gauge anomalies, so we do not consider them in what follows.

In addition to the SM Higgs doublet, $H$, the scalar sector is also augmented with a new $SU(2)_W^{}$ doublet scalar having hypercharge $1/2$ ($Y=1/2$), $\Phi$, and two SM gauge singlet scalars, $S_L$ and $S_{\mu\tau}$.
Here, $S_{\mu\tau}^{}$ breaks the $U(1)_{\lmlt}^{}$ symmetry and gives an MeV scale mass to the $U(1)_{\lmlt}^{}$ gauge boson, $\Zp$, after developing a vacuum expectation value (VEV).
As studied in Refs. \cite{Araki:2014ona,Araki:2015mya}, the $g_\mu -2$ problem and the dip in the IceCube data can simultaneously be solved with $M_{\Zp}^{}=11$ MeV and $g_{\mu\tau}^{} = 5 \times 10^{-4}$, where $M_{\Zp}^{}$ is a mass of $\Zp$ and $g_{\mu\tau}^{}$ is the gauge coupling constant of $U(1)_{\lmlt}^{}$.
We refer to these values in this work and require $S_{\mu\tau}^{}$ to develop
\begin{eqnarray}
\vev{S_{\mu\tau}^{}} = \frac{M_{\Zp}^{}}{g_{\mu\tau}^{}} 
\simeq 20 ~{\rm GeV} .
\label{eq:Vmt}
\end{eqnarray}

Furthermore, we introduce a global lepton number symmetry $U(1)_L$, which is explicitly broken in the scalar potential, see Sec. \ref{sec:scalar}.
The gauge singlet scalar $S_L$ and the $SU(2)_W$ doublet scalar $\Phi$ are introduced to spontaneously break $U(1)_L$ and to generate tiny masses for neutrinos.
In Table \ref{tab1}, we summarize the particle contents and the charge assignments of the models.

The Lagrangian relevant to lepton masses is given by 
\begin{eqnarray}
-{\cal L}_\ell^{}
= &&
  y_e^{} \overline{\ell_L^{}} H \ell_{R}^{}
+ y_D^{} \overline{\ell_L^{}}\tilde{H} N_R
+ y_N^{} \overline{\ell_L^{}}\tilde{\Phi} (N_L^{})^c 
+ y_S^{} \overline{(N_R^{})^{c}}(N_L^{})^c S_{\mu\tau}^{}
+ y_S^\prime \overline{(N_R^{})^{c}}(N_L^{})^c S_{\mu\tau}^{*}
\nonumber \\
&&
+ M_S^{} \overline{(N_R^{})^{c}}(N_L^{})^c
+ \frac{y_{RR}^{}}{2} \overline{(N_R^{})^{c}}N_R^{} S_L^{}
+ \frac{y_{LL}^{}}{2} \overline{N_L^{}}(N_L^{})^c S_L^{*}
+ h.c.~,
\label{eq:Lag}
\end{eqnarray}
where $\ell_L$ stands for the $SU(2)_W^{}$ doublet left-handed leptons, $\ell_R$ is the $SU(2)_W^{}$ singlet right-handed leptons, 
$\tilde{H}(\tilde{\Phi})=i\tau_2 H^*(\Phi^*)$ with the second Pauli matrix $\tau_2^{}$, and the flavor indices are omitted.
Note that $\overline{\ell_L}\tilde{H} (N_L)^c$ and 
$\overline{\ell_L}\tilde{\Phi} N_R$ are forbidden by $U(1)_L$, and the charged lepton Yukawa matrix, $y_e^{}$, is forced to be diagonal by $U(1)_{\lmlt}$.
After the scalars develop VEVs, the following Dirac and Majorana type neutrino mass matrices arise:
\begin{eqnarray}
&&
m_D^{} = y_D^{} \vev{H} ,~~~
m_N^{} = y_N^{} \vev{\Phi} ,~~~
m_S^{} = M_S^{} + (y_S^{}+y_S^\prime) \vev{S_{\mu\tau}^{}} ,
\nonumber \\
&&
m_{LL}^{} = y_{LL}^{} \vev{S_L^{}} ,~~~
m_{RR}^{} = y_{RR}^{} \vev{S_L^{}} ,
\label{eq:Ms}
\end{eqnarray}
and they take the forms of
\begin{eqnarray}
&&
m_D^{} = 
\bmx{cc}
m_{d}^{ee} & 0 \\
0 & m_{d}^{\mu\mu} \\
0 & 0
\emx ,
~~~
m_N^{} = 
\bmx{cc}
m_{n}^{ee} & 0 \\
0 & 0 \\
0 & m_{n}^{\tau\mu}
\emx  ,
~~~
m_S^{} = 
\bmx{cc}
m_{s}^{ee} & m_{s}^{e\mu} \\
m_{s}^{\mu e} & m_{s}^{\mu\mu}
\emx ,
\nonumber \\
&&
m_{LL}^{} = 
\bmx{cc}
m_{L}^{} & 0 \\
0 & 0 
\emx ,
~~~
m_{RR}^{} = 
\bmx{cc}
m_{R}^{} & 0 \\
0 & 0 
\emx ,
\end{eqnarray}
in the case of Model-A, or 
\begin{eqnarray}
&&
m_D^{} = 
\bmx{cc}
m_{d}^{ee} & 0 \\
0 & 0 \\
0 & m_{d}^{\tau\tau}
\emx ,
~~~
m_N^{} = 
\bmx{cc}
m_{n}^{ee} & 0 \\
0 & m_{n}^{\mu\tau} \\
0 & 0
\emx  ,
~~~
m_S^{} = 
\bmx{cc}
m_s^{ee} & m_s^{e\tau} \\
m_s^{\tau e} & m_s^{\tau\tau}
\emx ,
\nonumber \\
&&
m_{LL}^{} = 
\bmx{cc}
m_{L}^{} & 0 \\
0 & 0 
\emx ,
~~~
m_{RR}^{} = 
\bmx{cc}
m_{R}^{} & 0 \\
0 & 0 
\emx ,
\end{eqnarray}
in the case of Model-B.
The differences between Model-A and Model-B will be studied in more detail in Sec. \ref{sec:num}.

\section{Neutrino masses and mixing}
\label{sec:neutrino}
The mass matrices in Eq. (\ref{eq:Ms}) compose the $7\times 7$ neutrino mass matrix
\begin{eqnarray}
{\cal M}_\nu 
=
\bmx{ccc}
 0 & m_D^{} & m_N^{} \\
 (m_D^{})^T & m_{RR}^{} & m_S^{} \\
 (m_N^{})^T & (m_S^{})^T & m_{LL}^{}
\emx
=
\bmx{cc}
0 & {\cal M}_{DN}^{} \\
({\cal M}_{DN})^T & {\cal M}_{4\times 4}^{}
\emx ,
\end{eqnarray}
where 
\begin{eqnarray}
{\cal M}_{DN}^{} = (m_D^{}, ~m_N^{}),
~~~~
{\cal M}_{4\times 4}^{} =
\bmx{cc}
m_{RR}^{} & m_S^{} \\
(m_S^{})^T & m_{LL}^{}
\emx .
\end{eqnarray}
In order to generate tiny neutrino masses, we invoke the seesaw mechanism by assuming the hierarchy of 
\begin{eqnarray}
m_S^{} \gg m_D^{} \gg m_N^{}, m_{RR}^{}, m_{LL}^{}
~.
\label{eq:seesaw_hier}
\end{eqnarray}
Then, one can approximately block-diagonalize ${\cal M}_\nu$ as 
\begin{eqnarray}
U_2^T U_1^T
\bmx{cc}
0 & {\cal M}_{DN}^{} \\
({\cal M}_{DN})^T & {\cal M}_{4\times 4}^{}
\emx
U_1 U_2
\simeq
U_2^T
\bmx{cc}
m_\nu & 0 \\
0 & {\cal M}_{4\times 4}
\emx
U_2
\simeq
\bmx{ccc}
m_\nu & 0 & 0 \\
0 & M_1 & 0 \\
0 & 0 & M_2
\emx ,
\label{eq:blckdiag}
\end{eqnarray}
with
\begin{eqnarray}
U_1 =
\bmx{cc}
{\bf 1}_{3\times 3}-\frac{1}{2}\Theta^*\Theta^T & \Theta^* \\
-\Theta^T & {\bf 1}_{4\times 4}-\frac{1}{2}\Theta^T\Theta^*
\emx, ~~~
U_2 = 
\bmx{ccc}
{\bf 1}_{3\times 3} & 0 & 0 \\
0 & \frac{1}{\sqrt{2}} (V_S^{})^* & \frac{1}{\sqrt{2}} (V_S^{})^* \\
0 & -\frac{1}{\sqrt{2}} V_S^{} & \frac{1}{\sqrt{2}} V_S^{}
\emx ~,
\label{eq:U1U2}
\end{eqnarray}
where the active neutrino mass matrix is given by 
\begin{eqnarray}
m_\nu \simeq -\Theta ({\cal M}_{DN})^T,
\end{eqnarray}
the mixing parameter $\Theta$ is defined as 
\begin{eqnarray}
\Theta={\cal M}_{DN} ({\cal M}_{4\times 4})^{-1},
\end{eqnarray}
${\bf 1}_{n\times n}$ stands for the $n\times n$ unit matrix, and $V_S^{}$ is a diagonalizing matrix of $m_S^{}$.
Here, the inverse matrix of ${\cal M}_{4\times 4}$ is approximately given by
\begin{eqnarray}
({\cal M}_{4\times 4})^{-1}
\simeq
\bmx{cc}
-(m_S^T)^{-1} m_{LL}^{} (m_S^{})^{-1} & (m_S^T)^{-1} \\
(m_S)^{-1} & -(m_S^{})^{-1} m_{RR}^{} (m_S^{T})^{-1}
\emx ~,
\end{eqnarray}
which in turn leads to
\begin{eqnarray}
\Theta = \left(~ 
-m_D^{}(m_S^T)^{-1} m_{LL}^{} (m_S^{})^{-1} + m_N^{}(m_S^{})^{-1}, 
~~m_D^{}(m_S^T)^{-1} - m_N^{}(m_S^{})^{-1} m_{RR}^{} (m_S^{T})^{-1}
~\right)~.
\label{eq:Theta}
\end{eqnarray}
As a result, the active neutrino mass matrix is derived as
\begin{eqnarray}
m_\nu 
&\simeq & 
- m_N^{}(m_S^{})^{-1}m_D^T - m_D^{}(m_S^T)^{-1}m_N^T 
+ m_D^{}(m_S^T)^{-1} m_{LL}^{} (m_S^{})^{-1}m_D^T
~,
\label{eq:Mnu}
\end{eqnarray}
where we ignore $m_N^{}(m_S^{})^{-1} m_{RR}^{} (m_S^{T})^{-1}m_N^T$ due to Eq. (\ref{eq:seesaw_hier}).
We emphasize that the obtained mass matrix is twofold: the first two terms stem from the so-called linear seesaw mechanism, while the last term from the inverse seesaw one.
As will be discussed in Sec. \ref{sec:num}, in order to reproduce experimentally allowed neutrino mass spectrum and mixing, both contributions are necessary.
To this end, throughout this paper, we suppose the following parameter setting:
\begin{eqnarray}
&&
\vev{H} = 246 ~{\rm GeV},~~
\vev{\Phi} = {\cal O}(10^{-8}) ~{\rm GeV},~~
\vev{S_L^{}} = {\cal O}(10^{-7}) ~{\rm GeV},~~
\vev{S_{\mu\tau}^{}} = 20 ~{\rm GeV},
\label{eq:VEVs}
\\
&&
M_S = {\cal O}(10^2)~{\rm GeV} ,~~
y_D^{} = {\cal O}(10^{-2}) ,~~
y_N^{},~y_S^{},~y_{LL}^{},~y_{RR}^{} = {\cal O}(1) ,
\label{eq:ykws}
\end{eqnarray}
which results in $m_\nu^{} = {\cal O}(10^{-10})$ GeV.
We here note that $H$ serves as the SM like Higgs doublet, and $\vev{S_{\mu\tau}^{}}$ is determined to induce an MeV scale mass to $\Zp$ in Eq. (\ref{eq:Vmt}).
A more detailed discussion of the scalar potential is given in Sec. \ref{sec:scalar}.

Let us mention the active-sterile mixing and derive the leptonic mixing matrix in our framework.
The mixing between the active neutrinos and the sterile ones is described in Eq. (\ref{eq:U1U2}).
In view of the hierarchy Eq. (\ref{eq:seesaw_hier}), we simplify $\Theta$ in Eq. (\ref{eq:Theta}) as 
\begin{eqnarray}
\Theta 
\simeq \left(~ 0, ~~m_D^{}(m_S^T)^{-1} ~\right)
= (0,~\eta),
\end{eqnarray}
leading to
\begin{eqnarray}
\bmx{c}
(\nu_L)^c \\ N_R \\ (N_L)^c
\emx
=
\bmx{ccc}
{\bf 1}_{3\times 3}-\frac{1}{2}\eta^* \eta^T & 
-\frac{1}{\sqrt{2}}\eta^* V_S^{} &
 \frac{1}{\sqrt{2}}\eta^* V_S^{} 
\\
0 & \frac{1}{\sqrt{2}} (V_S^{})^* &
\frac{1}{\sqrt{2}} (V_S^{})^*
\\
-\eta^T & 
-\frac{1}{\sqrt{2}} \left[ {\bf 1}_{2\times 2}-\frac{1}{2}\eta^T\eta^* \right]V_S^{} &
\frac{1}{\sqrt{2}} \left[ {\bf 1}_{2\times 2}-\frac{1}{2}\eta^T\eta^* \right]V_S^{}
\emx
\bmx{c}
\nu^c \\ N_1 \\ N_2
\emx ,
\label{eq:AS-mix}
\end{eqnarray}
and the leptonic mixing matrix is obtained as 
\begin{eqnarray}
{\cal N} \simeq \left[ 1-\frac{1}{2}\eta\eta^\dag \right]V_{\rm MNS}^{}P .
\label{eq:Ulep}
\end{eqnarray}
Here, 
$P={\rm Diag}( 1, e^{i\frac{\alpha_{21}^{}}{2}}_{}, e^{i\frac{\alpha_{31}^{}}{2}}_{})$
includes the Majorana CP-violating phases, and $V_{\rm MNS}^{}$ is the MNS mixing matrix \cite{Maki:1962mu} parametrized as 
\begin{eqnarray}
V_{\rm MNS}^{} = 
\bmx{ccc}
 c_{12}c_{13} & s_{12}c_{13} & s_{13}e^{-i\delta} \\
-s_{12}c_{23}-c_{12}s_{23}s_{13}e^{i\delta} & 
 c_{12}c_{23}-s_{12}s_{23}s_{13}e^{i\delta} &
 s_{23}c_{13} \\
 s_{12}s_{23}-c_{12}c_{23}s_{13}e^{i\delta} & 
-c_{12}s_{23}-s_{12}c_{23}s_{13}e^{i\delta} &
 c_{23}c_{13} 
\emx ,
\end{eqnarray}
where $c_{ij}=\cos\theta_{ij}$, $s_{ij}=\sin\theta_{ij}$ and $\delta$ denotes the Dirac CP-violating phase.
In Eq. (\ref{eq:Ulep}), the full mixing matrix ${\cal N}$ is no longer a unitary matrix due to the presence of the second term, and values of $\eta$'s are, at present, moderately constrained by flavor and electroweak precision experiments.
We use these constraints in numerically calculating neutrino masses and mixing in the next section.

\section{Numerical calculations}
\label{sec:num}
We here numerically diagonalize the active neutrino mass matrix and check its consistency with experiments.
In the calculations, we place the $3\sigma$ constraints on the neutrino oscillation parameters obtained in Ref. \cite{Esteban:2018azc}:
\begin{eqnarray}
&&
\sin^2\theta_{12}=0.275 - 0.350 ~~(0.275 - 0.350),
\nonumber \\
&&
\sin^2\theta_{23}=0.418 - 0.627 ~~(0.423 - 0.629),
\nonumber \\
&&
\sin^2\theta_{13}=0.02045 - 0.02439 ~~(0.02068 - 0.02463),
\nonumber \\
&&
\Delta m_{21}^2/10^{-5} = 6.79 - 8.01 ~~(6.79 - 8.01),
\nonumber \\
&&
\Delta m_{31}^2/10^{-3} = 2.427 - 2.625 ~~(\Delta m_{23}^2/10^{-3} = 2.412 - 2.611),
\nonumber \\
&&
\delta = 125^\circ - 392^\circ ~~(196^\circ - 360^\circ)~,
\label{eq:gfit-3s}
\end{eqnarray}
for normal (inverted) mass ordering, and the $2\sigma$ upper bounds on $\eta$'s derived in Ref. \cite{Fernandez-Martinez:2016lgt}:
\begin{eqnarray}
&&
(\eta\eta^\dag)_{ee}^{} < 2.5 \times 10^{-3} ,~~
(\eta\eta^\dag)_{\mu\mu}^{} < 4.4 \times 10^{-4} ,~~
(\eta\eta^\dag)_{\tau\tau}^{} < 5.6 \times 10^{-3} ,
\nonumber \\
&&
(\eta\eta^\dag)_{e\mu}^{} < 2.4 \times 10^{-5} ,~~
(\eta\eta^\dag)_{e\tau}^{} < 2.7 \times 10^{-3} ,~~
(\eta\eta^\dag)_{\mu\tau}^{} < 1.2 \times 10^{-3} .
\label{eq:nonU-2s}
\end{eqnarray}

In terms of $\eta$, the active neutrino mass matrix in Eq. (\ref{eq:Mnu}) is rewritten by
\begin{eqnarray}
m_\nu^{} = - m_N^{}\eta_{}^T - \eta m_N^T + \eta m_{LL}^{}\eta_{}^T ,
\end{eqnarray}
and it takes the form of
\begin{eqnarray}
m_\nu^{} 
= -
\bmx{ccc}
2m_n^{ee}\eta^{ee} & m_n^{ee}\eta^{\mu e} & m_n^{\tau\mu}\eta^{e\mu} \\
m_n^{ee}\eta^{\mu e} & 0 & m_n^{\tau\mu}\eta^{\mu\mu} \\
m_n^{\tau\mu}\eta^{e\mu} & m_n^{\tau\mu}\eta^{\mu\mu} & 0
\emx
+ m_L
\bmx{ccc}
\eta^{ee}\eta^{ee} & \eta^{ee}\eta^{\mu e} & 0\\
\eta^{\mu e}\eta^{ee} & \eta^{\mu e}\eta^{\mu e} & 0 \\
0 & 0 & 0 \\
\emx~,
\label{eq:m33}
\end{eqnarray}
for Model-A, while it takes 
\begin{eqnarray}
m_\nu^{} 
= -
\bmx{ccc}
2m_n^{ee}\eta^{ee} & m_n^{\mu\tau}\eta^{e\tau} & m_n^{ee}\eta^{\tau e} \\
m_n^{\mu\tau}\eta^{e\tau} & 0 & m_n^{\mu\tau}\eta^{\tau\tau} \\
m_n^{ee}\eta^{\tau e} & m_n^{\mu\tau}\eta^{\tau\tau} & 0
\emx
+ m_L
\bmx{ccc}
\eta^{ee}\eta^{ee} & 0 & \eta^{ee}\eta^{\tau e} \\
0 & 0 & 0 \\
\eta^{\tau e}\eta^{ee} & 0 & \eta^{\tau e}\eta^{\tau e}
\emx ,
\label{eq:m22}
\end{eqnarray}
for Model-B.
As can be seen, both the neutrino mass matrices contain one zero\footnote{
There is another contribution to the neutrino mass term from one-loop diagrams \cite{Dev:2012sg}.
In our framework, the diagrams yield a new contribution, which is proportional to $m_D^{}m_{RR}^{}m_D^T$, to $(m_\nu^{})_{ee}^{}$ and do not break the obtained one-zero textures.
Thus, by tuning $y_{RR}^{}$ in Eq. (\ref{eq:Lag}), we simply assume that the contribution is negligibly small compared with Eqs. (\ref{eq:m33}) and (\ref{eq:m22}).
}, 
and thus one can predict two observables.
In the following subsections, we calculate the effective mass of neutrinoless double beta decay, $\vev{m_{ee}^{}}$, as a function of the sum of active neutrino masses, $\sum_i m_i$, and check the consistency with the current bounds:
\begin{eqnarray}
\vev{m_{ee}^{}} < 0.061 - 0.165~{\rm eV} ,
\label{eq:Kam}
\end{eqnarray}
from the KamLAND-Zen Collaboration \cite{KamLAND-Zen:2016pfg}, where the uncertainty comes from the nuclear matrix element calculation, and the combined upper bound
\begin{eqnarray}
\sum_{i=1}^{3} m_i^{} < 0.12 ~{\rm eV} ,
\label{eq:Pla}
\end{eqnarray}
from the Planck Collaboration \cite{Aghanim:2018eyx}.

\subsection{Inverted ordering}
\begin{figure}[h]
\begin{flushleft}
\includegraphics[width=7.2cm]{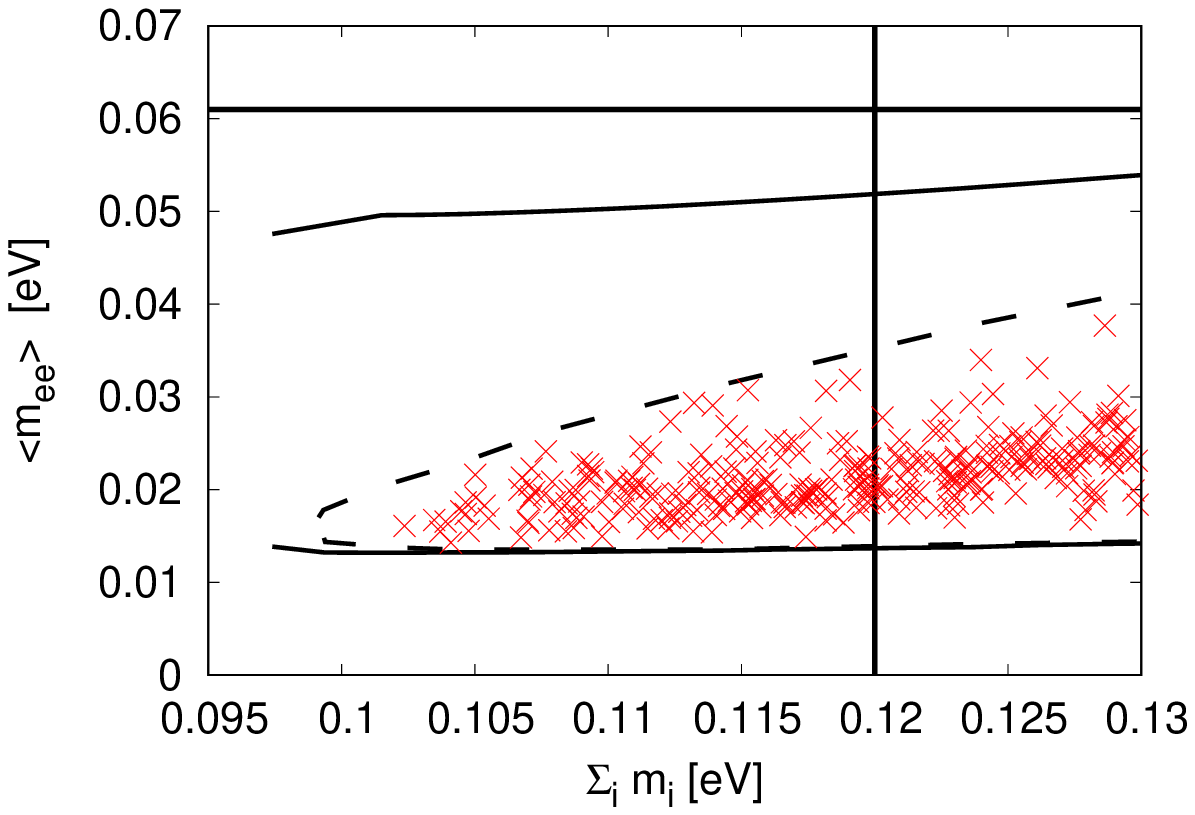}
\hspace{0.5cm}
\includegraphics[width=7.2cm]{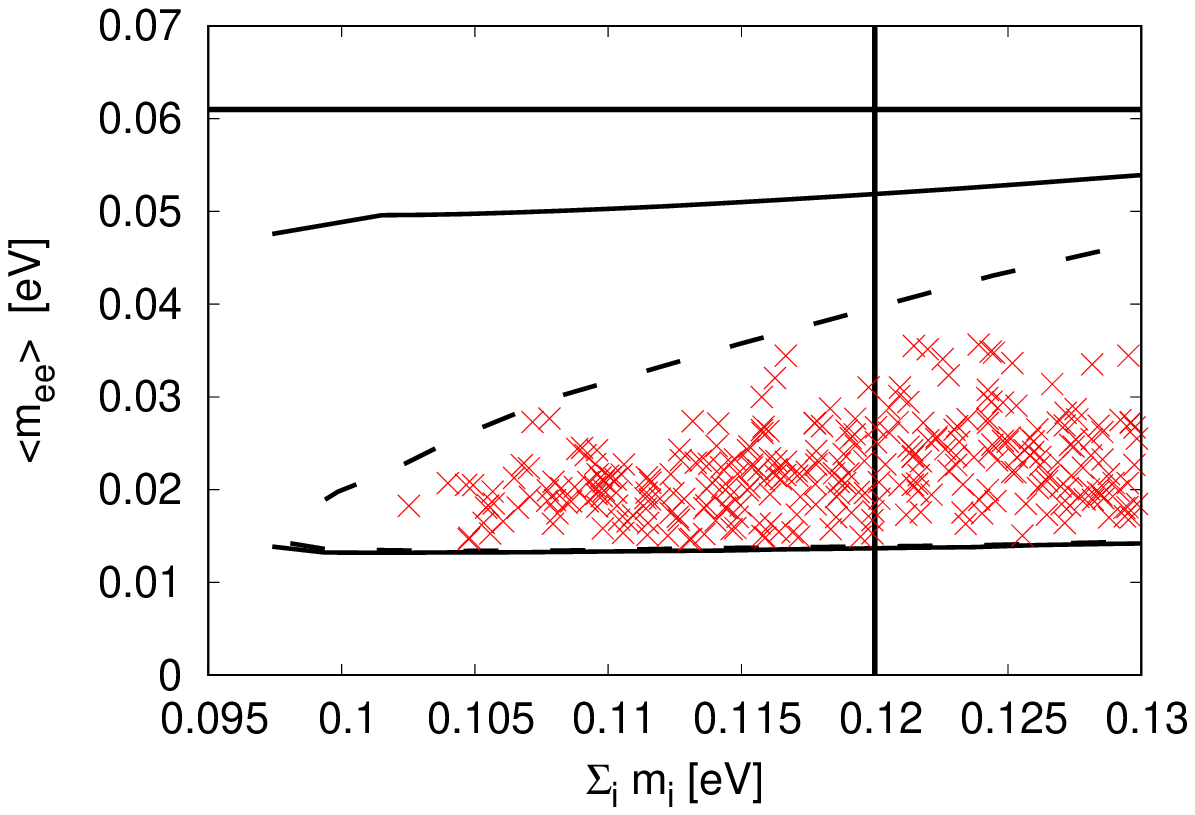}
\end{flushleft}
\vspace{-1.5cm}
\caption{
The effective mass of neutrinoless double beta decay, $\vev{m_{ee}}$, as a function of the sum of active neutrino masses, $\sum_i m_i$, in the case of inverted mass ordering, for Model-A (left panel) and Model-B (right panel).
The solid curves display $3\sigma$ upper and lower bounds corresponding to Eq. (\ref{eq:gfit-3s}).
The regions surrounded by the dashed curves are full parameter regions of the models, while the red crosses, $\times$, represent predicted points when varying the parameters as in Eq. (\ref{eq:para}).
The horizontal solid lines indicate the tightest upper bound of Eq. (\ref{eq:Kam}), and the vertical lines indicate Eq. (\ref{eq:Pla}).
}
\label{fig:inverted}
\end{figure}
Let us first investigate the inverted ordering case.
In view of our parameter setting given in Eqs. (\ref{eq:VEVs}) and (\ref{eq:ykws}), we vary the active neutrino mass matrix within the following ranges:
\begin{eqnarray}
|\eta_{}^{\alpha\beta}| = 0.001 - 0.02,~~~
|m_n^{\alpha\beta}| = 10^{-9} - 10^{-8}~{\rm GeV},~~~
|m_L^{}| = 10^{-8} -10^{-7}~{\rm GeV} ,
\label{eq:para}
\end{eqnarray}
while imposing the constraints in Eqs. (\ref{eq:gfit-3s}) and (\ref{eq:nonU-2s}).
Here, we take a basis in which only $\eta^{ee}$ and $\eta^{e\mu}(\eta^{e\tau})$ are complex and vary their phases within $0-2\pi$.
We vary these parameters randomly with flat probability distributions in each range.
In Fig. \ref{fig:inverted}, we plot $\vev{m_{ee}^{}}$ as a function of $\sum_i m_i$ and find that there exist parameter regions (red crosses, $\times$) in which all the constraints can be satisfied for both Model-A and Model-B.
Note that, in the figures, the density of points has no statistical meanings; it shows the difficulty of finding solutions.
For instance, in the areas where the density is low, it is difficult to find solutions because relatively strong parameter tuning is necessary to satisfy Eqs. (\ref{eq:gfit-3s}) and (\ref{eq:nonU-2s}), especially the small squared-mass-differences.
As a reference, we also show the current $3\sigma$ upper and lower bounds (solid curves), which are derived by calculating
\begin{eqnarray}
\vev{m_{ee}^{}} =
| (c_{12}c_{13})^2 m_1^{}
+ (s_{12}c_{13})^2 m_2^{}e^{i\alpha_{21}^{}}
+ (s_{13}e^{-i\delta})^2 m_3^{}e^{i\alpha_{31}^{}} | ,
\end{eqnarray}
with Eq. (\ref{eq:gfit-3s}) while varying the Majorana phases within $0-2\pi$. Also, full parameter regions of the models (dashed curves) are shown by solving the condition $(m_\nu^{})_{\tau\tau}^{}=0$:
\begin{eqnarray}
 (s_{12}s_{23}-c_{12}c_{23}s_{13}e^{i\delta})^2 m_1^{}
+(c_{12}s_{23}+s_{12}c_{23}s_{13}e^{i\delta})^2 m_2^{}e_{}^{i\alpha_{21}^{}}
+(c_{23}c_{13})^2 m_3^{}e_{}^{i\alpha_{31}^{}}
=0 ,
\label{eq:m33=0}
\end{eqnarray}
for Model-A, or $(m_\nu^{})_{\mu\mu}^{}=0$:
\begin{eqnarray}
 (s_{12}c_{23}+c_{12}s_{23}s_{13}e^{i\delta})^2 m_1^{}
+(c_{12}c_{23}-s_{12}s_{23}s_{13}e^{i\delta})^2 m_2^{}e_{}^{\alpha_{21}^{}}
+(s_{23}c_{13})^2 m_3^{}e_{}^{\alpha_{31}^{}}
=0 ,
\label{eq:m22=0}
\end{eqnarray}
for Model-B.
Unfortunately, the current sensitivity on $\vev{m_{ee}^{}}$ is not enough to test the predicted regions.
However, in next generation experiments, the sensitivity is hoped to reach $\vev{m_{ee}^{}}={\cal O}(0.01)$ eV \cite{Next}, so our models would be tested in the near future.

\subsection{Normal ordering}
\begin{figure}[h]
\begin{flushleft}
\includegraphics[width=7.2cm]{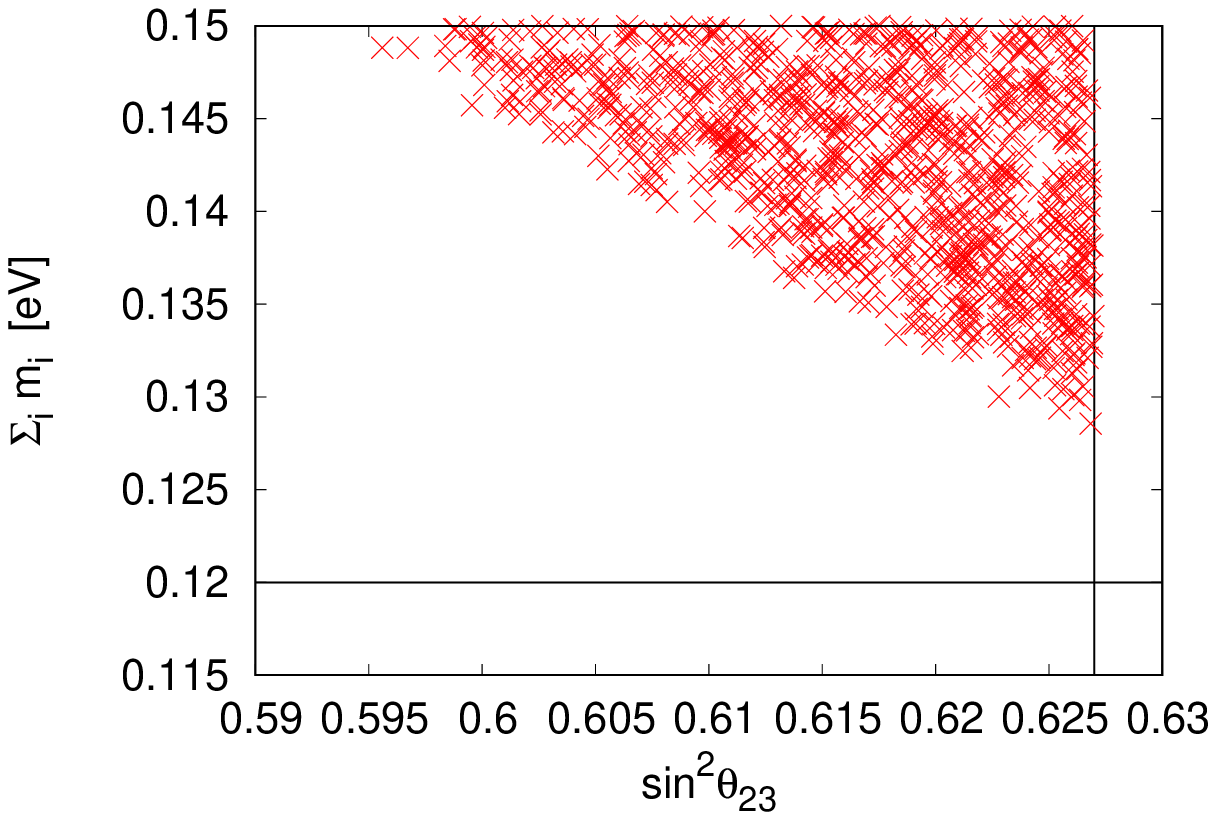}
\hspace{0.5cm}
\includegraphics[width=7.2cm]{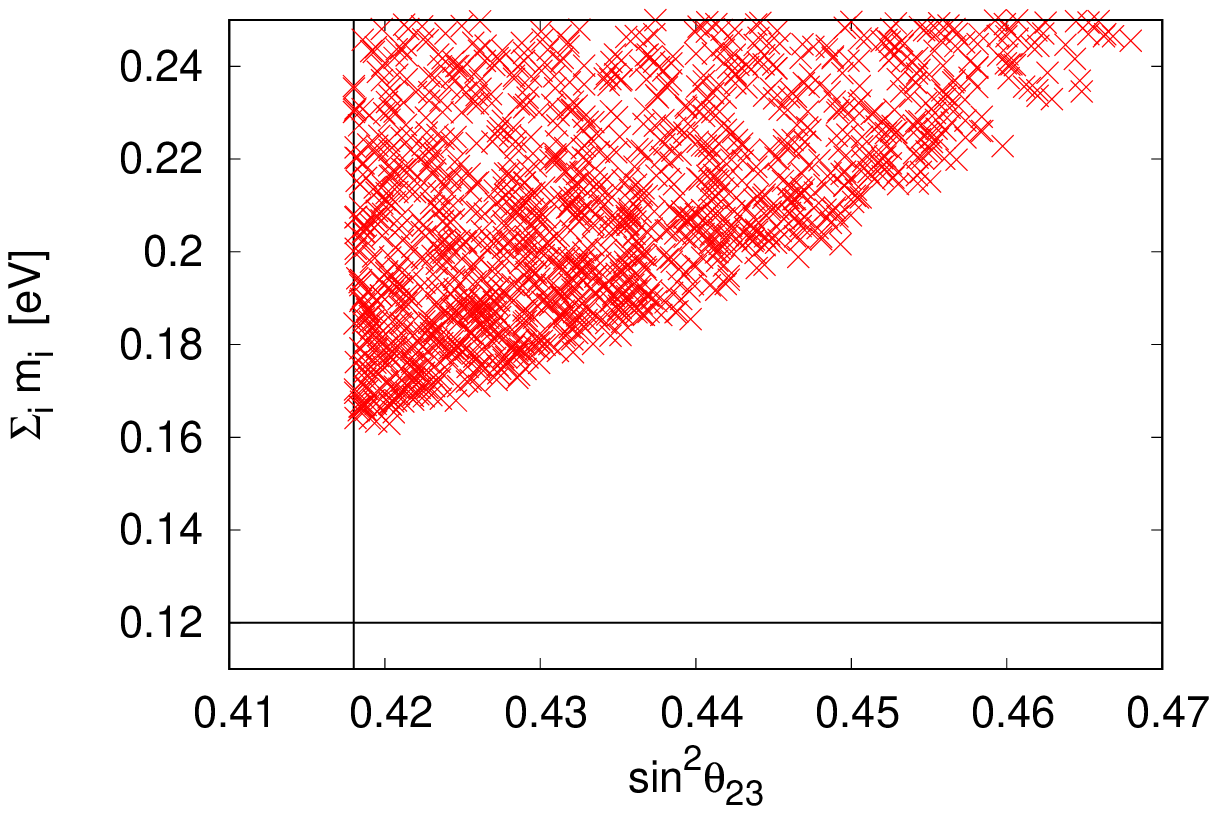}
\end{flushleft}
\vspace{-1.5cm}
\caption{
The sum of active neutrino masses, $\sum_i m_i$, as a function of $\sin^2\theta_{23}$ in the case of normal mass ordering, for Model-A (left panel) and Model-B (right panel).
The horizontal solid lines correspond to Eq. (\ref{eq:Pla}), while the vertical lines are $\sin^2\theta_{23}=0.627$ (left panel) and $\sin^2\theta_{23}=0.418$ (right panel) which are the upper and the lower bound in Eq. (\ref{eq:gfit-3s}), respectively.
}
\label{fig:normal}
\end{figure}
For normal mass ordering, we find that both of the one-zero textures are now excluded by the Planck bound.
In order to show this conclusion, instead of investigating the mass matrices in Eqs. (\ref{eq:m33}) and (\ref{eq:m22}), we calculate $\sum_i m_i$ by solving Eqs. (\ref{eq:m33=0}) and (\ref{eq:m22=0}) and check their consistency.
In Fig. \ref{fig:normal}, we show $\sum_i m_i$ as a function of $\sin^2\theta_{23}$ for the cases of $(m_\nu^{})_{\tau\tau}^{}=0$ (left panel) and $(m_\nu^{})_{\mu\mu}^{}=0$ (right panel).
The neutrino oscillation parameters are randomly scattered within the $3\sigma$ ranges given in Eq. (\ref{eq:gfit-3s}), while the Majorana phases are varied within $0-2\pi$.
As can be seen from the figures, $\sum_i m_i$ is constrained to be $\gtrsim 0.13$ eV ($\gtrsim 0.16$ eV) by the upper (lower) bound of $\sin^2\theta_{23}$ in the case of $(m_\nu^{})_{\tau\tau}^{}=0$ ($(m_\nu^{})_{\mu\mu}^{}=0$), which is clearly inconsistent with Eq. (\ref{eq:Pla}).


\section{Scalar sector}
\label{sec:scalar}
The scalar potential invariant under ${\cal G}_{\rm SM} \times U(1)_{\lmlt}^{} \times U(1)_L^{}$, where ${\cal G}_{\rm SM}^{}$ denotes the SM gauge group, is given by
\begin{eqnarray}
V &=&
  m_H^2 |H|^2 + m_\Phi^2|\Phi|^2 
+ m_L^2|S_L^{}|^2 + m_{\mu\tau}^2|S_{\mu\tau}^{}|^2
\nonumber \\
&&~~
+ \mu[H^\dag \Phi S_L^\dag + h.c.] 
\nonumber \\
&&~~
+ \lambda_1 |H|^4 + \lambda_2 |\Phi|^4 
+ \lambda_3 |H|^2|\Phi|^2 + \lambda_4(H^\dag\Phi)(\Phi^\dag H) 
\nonumber \\
&&~~
+ \lambda_5 |S_{\mu\tau}|^4 + \lambda_6 |H|^2|S_{\mu\tau}|^2 
+ \lambda_7 |\Phi|^2|S_{\mu\tau}|^2
\nonumber \\
&&~~
+ \lambda_8 |S_L|^4 + \lambda_9 |H|^2|S_L|^2 
+ \lambda_{10} |\Phi|^2|S_L|^2 
+ \lambda_{11}|S_L|^2 |S_{\mu\tau}|^2
~.
\label{eq:Vpot}
\end{eqnarray}
Since the potential preserves the global $U(1)_L$ symmetry, a massless Nambu-Goldstone boson (NGB) appears after the spontaneous symmetry breaking.
In addition, as will be seen later, an almost massless scalar boson appears in the CP-even sector for our parameter setting given in Eq. (\ref{eq:VEVs}).
The existence of the NGB and the almost massless scalar boson has interesting implications for observations of high energy cosmic neutrinos.
In Sec. \ref{sec:SnI}, we show that if these particles have MeV scale masses, they would significantly attenuate the flux of high energy cosmic neutrinos.
In order to give finite masses to them, we here introduce the following $U(1)_L^{}$ breaking term:
\begin{eqnarray}
V_{\not{L}} = B^2[ H^\dag \Phi + h.c. ]~.
\label{eq:Vlb}
\end{eqnarray} 
This breaking term becomes important also when we solve stationary conditions of the potential: tiny VEVs can naturally be obtained for $|m_\Phi^2| \gg |B^2|$ \cite{Ma:2000cc,Haba:2011ra}.
Note that in the potential all the parameters are taken to be real.

We define components of the scalars and their VEVs as
\begin{eqnarray}
&&
H=\bmx{c} h^+ \\ \frac{1}{\sqrt{2}}(v_{\rm ew}^{} + h + i\eta) \emx ,~~
\Phi=\bmx{c} \phi^+ \\ \frac{1}{\sqrt{2}}(v_\Phi + \phi + i\xi) \emx ,
\nonumber \\
&&
S_L^{} = \frac{1}{\sqrt{2}}(v_L^{} + s_L^{} + i\sigma_L^{}) ,~~
S_{\mu\tau}^{} = \frac{1}{\sqrt{2}}(v_{\mu\tau}^{} + s_{\mu\tau}^{} + i\sigma_{\mu\tau}^{}) ~.
\end{eqnarray}
Then, the stationary conditions are obtained as 
\begin{eqnarray}
v_{\rm ew}\left[ 
m_H^2 + \lambda_1 v_{\rm ew}^2 + \frac{\lambda_3+\lambda_4}{2}v_\Phi^2 + \frac{\lambda_6}{2} v_{\mu\tau}^2 + \frac{\lambda_9}{2} v_L^2
\right] + \frac{\mu}{\sqrt{2}} v_\Phi v_L + B^2 v_\Phi
&=& 0 ~,
\nonumber \\
v_{\Phi}\left[ 
m_\Phi^2 + \lambda_2 v_{\Phi}^2 + \frac{\lambda_3+\lambda_4}{2} v_{\rm ew}^2 + \frac{\lambda_7}{2} v_{\mu\tau}^2 + \frac{\lambda_{10}}{2} v_L^2
\right] + \frac{\mu}{\sqrt{2}} v_{\rm ew} v_L + B^2 v_{\rm ew}
&=& 0 ~,
\nonumber \\
v_L\left[ 
m_L^2 + \lambda_8 v_L^2 + \frac{\lambda_9}{2} v_{\rm ew}^2 + \frac{\lambda_{10}}{2} v_\Phi^2 + \frac{\lambda_{11}}{2} v_{\mu\tau}^2
\right] 
+ \frac{\mu}{\sqrt{2}} v_{\rm ew} v_\Phi
&=& 0 ~,
\nonumber \\
m_{\mu\tau}^2 + \lambda_5 v_{\mu\tau}^2 + \frac{\lambda_6}{2} v_{\rm ew}^2 + \frac{\lambda_7}{2} v_\Phi^2 + \frac{\lambda_{11}}{2} v_L^2 &=& 0 ~.
\end{eqnarray}
In the second and the third condition, for $|m_\Phi^2|,~|m_L^2| \gg v_{\rm ew}^2,~v_{\mu\tau}^2,~v_L^2,~v_\Phi^2$, the conditions can be written as
\begin{eqnarray}
v_L \simeq -\frac{\mu}{\sqrt{2}}\frac{v_{\rm ew}}{m_L^2}v_\Phi ,~~
v_\Phi \simeq -\frac{B^2 v_{\rm ew}}{m_\Phi^2 - \frac{\mu^2}{2}\frac{v_{\rm ew}^2}{m_L^2}}.
\end{eqnarray}
Therefore, $v_\Phi$ and $v_L$ can simultaneously be small for $|m_L^2| \simeq |\mu v_{\rm ew}|$ and $|m_\Phi^2| \gg |B^2|({\rm or}~|\mu v_{\rm ew}|\gg |B^2|)$.
For instance, $v_\Phi={\cal O}(10^{-8})$ GeV and $v_L={\cal O}(10^{-7})$ GeV can be realized for $|m_\Phi^2|={\cal O}(10^{9})~{\rm GeV}^2$, $|m_L^2|={\cal O}(10^6)~{\rm GeV}^2$, $|\mu|={\cal O}(10^5)$ GeV and $|B^2|={\cal O}(10^{-1})~{\rm GeV}^2$.
In cases where $|B^2| \ll 10^{-6}~{\rm GeV}^2$, however, behavior of the stationary conditions changes; in such cases, parameter tuning is necessary to derive $v_L={\cal O}(10^{-7})$ GeV, $v_\Phi = {\cal O}(10^{-8})$ GeV, and hence $v_{\rm ew}/v_\Phi = {\cal O}(10^{10})$.

\subsection{Masses of charged scalars}
In a basis of $(h^\pm, \phi^\pm)$, the squared mass matrix of the charged components is given by
\begin{eqnarray}
M^2_{\pm} = 
\left( 
\frac{\mu}{\sqrt{2}}v_L^{} + B^2 + \frac{\lambda_4}{2}v_{\rm ew}^{}v_\Phi^{}
\right)
\bmx{cc}
 -\frac{v_\Phi^{}}{v_{\rm ew}^{}} & 1 \\
 1 & -\frac{v_{\rm ew}^{}}{v_\Phi^{}}
\emx~.
\end{eqnarray}
This mass matrix can easily be diagonalized, and one obtains an eigenvalue of
\begin{eqnarray}
m^2_{H^\pm} = 
-\left( 
\frac{\mu}{\sqrt{2}}v_L^{} + B^2 + \frac{\lambda_4}{2}v_{\rm ew}^{}v_\Phi^{}
\right)
\left( \frac{v_\Phi^{}}{v_{\rm ew}^{}}+\frac{v_{\rm ew}^{}}{v_\Phi^{}} 
\right) 
~,
\label{eq:Mpm}
\end{eqnarray} 
and a zero eigenvalue corresponding to the NGB for the SM $W^\pm$ gauge boson.

Charged scalar bosons are searched at the LHC experiments; they can be produced via top quark decays, $t \rightarrow H^\pm b$, or in association with a top and a bottom quark, $pp \rightarrow H^\pm tb$.
In our models, however, interactions between $H^\pm$ and quarks are suppressed by the small $v_\Phi$ (or large $\tan\beta=v_{\rm ew} / v_\Phi$), so the LHC constraints are irrelevant for our studies.
In contrast, at the LEP experiments, charged scalar bosons can be produced via the Drell-Yan process $e^+ e^- \rightarrow \gamma/Z \rightarrow H^+H^-$, which does not rely on interactions with quarks and may be applicable to our models.
In Ref. \cite{Abbiendi:2013hk}, lower bounds on a charged scalar boson mass are derived for several decay channels, in the framework of two-Higgs-doublet-models.
Since decays to quarks are strongly suppressed in our models, we refer to the bound for type-II two-Higgs-doublet-models with ${\rm Br}(H^\pm \rightarrow \tau^\pm \nu)=1$, i.e., $m_{H^\pm}^{} > 94$ GeV, and that for type-I models with ${\rm Br}(H^\pm \rightarrow W^\pm A) \simeq 1$ ($\tan\beta=100$), i.e., $m_{H^\pm}^{} > 84 - 89$ GeV, where $A$ denotes a lighter CP-odd neutral boson.
Also, in order to prevent $m_{H^\pm}^{}$ from being tachyonic, we require 
$|\frac{\mu}{\sqrt{2}}v_L^{} + B^2| > |\frac{\lambda_4}{2}v_{\rm ew}v_\Phi|$ in Eq. (\ref{eq:Mpm}).
As a consequence, we place a constraint of
\begin{eqnarray}
\left| 
\frac{\mu}{\sqrt{2}}v_L^{} + B^2
\right| 
> {\cal O}(10^{-6})~{\rm GeV}^2 
,
\end{eqnarray}
in what follows.

\subsection{Masses of CP-even neutral scalars}
\label{sec:even}
In a basis of $(h,~\phi,~s_L,~s_{\mu\tau})$, the squared mass matrix of the CP-even neutral components is given by
\begin{eqnarray}
M_{\rm even}^2 =
\bmx{cccc}
2\lambda_1 v_{\rm ew}^2 - \frac{\mu}{\sqrt{2}} \frac{v_\Phi v_L}{v_{\rm ew}} - B^2\frac{v_\Phi}{v_{\rm ew}} &  
(\lambda_3+\lambda_4)v_{\rm ew}v_\Phi + \frac{\mu}{\sqrt{2}} v_L + B^2 &  
\lambda_9 v_{\rm ew}v_L + \frac{\mu}{\sqrt{2}} v_\Phi &
\lambda_6 v_{\rm ew} v_{\mu\tau} 
\\
(\lambda_3+\lambda_4)v_{\rm ew}v_\Phi + \frac{\mu}{\sqrt{2}} v_L + B^2 &
2\lambda_2 v_\Phi^2 - \frac{\mu}{\sqrt{2}} \frac{v_L}{v_\Phi}v_{\rm ew} - B^2\frac{v_{\rm ew}}{v_\Phi} &
\lambda_{10}v_\Phi v_L + \frac{\mu}{\sqrt{2}} v_{\rm ew} &
\lambda_7 v_\Phi v_{\mu\tau}
\\
\lambda_9 v_{\rm ew}v_L + \frac{\mu}{\sqrt{2}} v_\Phi &
\lambda_{10}v_\Phi v_L + \frac{\mu}{\sqrt{2}} v_{\rm ew} &
2\lambda_8 v_L^2 - \frac{\mu}{\sqrt{2}}\frac{v_\Phi}{v_L}v_{\rm ew} &
\lambda_{11}v_{\mu\tau}v_L 
\\
\lambda_6 v_{\rm ew} v_{\mu\tau} & 
\lambda_7 v_\Phi v_{\mu\tau} & 
\lambda_{11}v_{\mu\tau}v_L	 & 
2\lambda_5 v_{\mu\tau}^2 \\
\emx~.
\end{eqnarray}
Although the mass matrix is too complicated to diagonalize in a general way, one can see that the $(\phi,~s_L)$ block part is dominated by the terms proportional to $\mu$ and an almost rank one matrix in case of $B^2=0$.
Consequently, an almost massless scalar boson inevitably appears in the absence of explicit $U(1)_L$ symmetry breaking for our parameter setting proposed in Eq. (\ref{eq:VEVs}).

In order to simplify the diagonalization, in this work, we restrict ourselves to the parameter region of
$|\frac{\mu}{\sqrt{2}}v_L^{} + B^2| < {\cal O}(1)~{\rm GeV}^2$
so that the mass matrix can be simplified to
\begin{eqnarray}
M_{\rm even}^2 
&\simeq&
\bmx{cccc}
2\lambda_1 v_{\rm ew}^2 & 0 & 0 & \lambda_6 v_{\rm ew} v_{\mu\tau} 
\\
0 & -\left( \frac{\mu}{\sqrt{2}}v_L + B^2\right)\frac{v_{\rm ew}}{v_\Phi} &  
\frac{\mu}{\sqrt{2}} v_{\rm ew} & 0  
\\
0 & \frac{\mu}{\sqrt{2}} v_{\rm ew} & 
-\frac{\mu}{\sqrt{2}}\frac{v_\Phi}{v_L}v_{\rm ew} & 0
\\
\lambda_6 v_{\rm ew} v_{\mu\tau} & 0 & 0 & 2\lambda_5 v_{\mu\tau}^2 \\
\emx ~.
\label{eq:Meven}
\end{eqnarray}
Note that $\mu$ and $B^2$ are also constrained by the lower bound on the charged scalar boson mass: $|\frac{\mu}{\sqrt{2}}v_L^{} + B^2| > {\cal O}(10^{-6})~{\rm GeV}^2$.
Then, the diagonalization can easily be done by rotating
\begin{eqnarray}
\bmx{c} h \\ \phi \\ s_L^{} \\ s_{\mu\tau}^{} \emx
\rightarrow
\bmx{cccc}
 c_1 & 0 & 0 & s_1 \\
 0 & 1 & 0 & 0 \\
 0 & 0 & 1 & 0 \\
-s_1 & 0 & 0 & c_1
\emx
\bmx{cccc}
1 & 0 & 0 & 0 \\
0 & c_2 & s_2 & 0 \\
0 & -s_2 & c_2 & 0 \\
0 & 0 & 0 & 1
\emx
\bmx{c} h_1 \\ h_2 \\ h_3 \\ h_4 \emx~,
\label{eq:even-mix}
\end{eqnarray}
where $c_{1(2)} = \cos\theta_{1(2)}$ and $s_{1(2)} = \sin\theta_{1(2)}$.
However, the exact expressions for the mixing angles and the eigenvalues are somewhat complicated.
Thus, we refrain from showing them and, instead, derive approximate expressions by considering certain limits.

Let us first consider the $(h,s_{\mu\tau})$ block part.
As we will show later, the value of $\lambda_6$ is restricted to be $\lambda_6 < {\cal O}(0.01)$ to suppress invisible decays of the Higgs boson.
Hence, given $v_{\rm ew}=246$ GeV and $v_{\mu\tau}=20$ GeV, one finds a hierarchy of 
$(M_{\rm even}^2)_{11} \gg (M_{\rm even}^2)_{44} \gg (M_{\rm even}^2)_{14}$ for $\lambda_{1,5}={\cal O}(1)$.
In this case, the diagonalization can approximately be done by
\begin{eqnarray}
s_1 \simeq -\frac{\lambda_6 v_{\mu\tau}}{2\lambda_1 v_{\rm ew}},
~~~~
c_1 \simeq 1-\frac{1}{2}
\left( 
\frac{\lambda_6 v_{\mu\tau}}{2\lambda_1 v_{\rm ew}} 
\right)^2~,
\label{eq:s1c1}
\end{eqnarray}
which lead to
\begin{eqnarray}
m_{h_1}^2 \simeq 2\lambda_1 v_{\rm ew}^2,
~~~
m_{h_4}^2 \simeq 2\lambda_5 v_{\mu\tau}^2.
\label{eq:h1h4}
\end{eqnarray}
The mass of $h_1^{}$ is proportional to $v_{\rm ew}^{}\simeq 246$ GeV, and thus we identify $h_1^{}$ as the SM Higgs boson, while the mass of $h_4$ is expected to be somewhat small, that is $m_{h_4}={\cal O}(10)$ GeV.

Next, we tackle with the $(\phi,s_L)$ block part for the following two cases.
\begin{itemize}
\item $|\frac{\mu}{\sqrt{2}}v_L| \ll |B^2|$\\
By utilizing a hierarchy of 
$(M_{\rm even}^2)_{22} \gg (M_{\rm even}^2)_{23} > (M_{\rm even}^2)_{33}$, the mixing $c_2$ and $s_2$ are derived as 
\begin{eqnarray}
s_2 \simeq \frac{\mu v_\Phi}{\sqrt{2} B^2},
~~~~
c_2 \simeq 1-\frac{1}{2}\left( \frac{\mu v_\Phi}{\sqrt{2} B^2} \right)^2 ,
\end{eqnarray}
which result in 
\begin{eqnarray}
&&m_{h_2}^2 
\simeq 
-\left( \frac{\mu}{\sqrt{2}} v_L + \sqrt{2}B^2\right)\frac{v_{\rm ew}}{v_\Phi}
\simeq 
-\sqrt{2}B^2 \frac{v_{\rm ew}}{v_\Phi},
\nonumber \\
~~~
&&m_{h_3}^2 
\simeq 
-\frac{\mu}{\sqrt{2}}\frac{v_\Phi}{v_L}v_{\rm ew}
+\frac{\mu^2 v_{\rm ew}v_\Phi}{\sqrt{2}(\mu v_L + \sqrt{2}B^2)}
\simeq 
-\frac{\mu}{\sqrt{2}}\frac{v_\Phi}{v_L}v_{\rm ew}.
\end{eqnarray}
Given $|\frac{\mu}{\sqrt{2}}v_L + B^2|\simeq|B^2|$ and thus 
${\cal O}(10^{-6})~{\rm GeV}^2 < |B^2| < {\cal O}(1)~{\rm GeV}^2 $, one obtains ${\cal O}(10^{2})~{\rm GeV} < m_{h_2} < {\cal O}(10^{5})~{\rm GeV}$.
In contrast, the mass of $h_3$ is determined by $\mu$; it is roughly given by $m_{h_3}^2 \simeq -\mu \times 10~{\rm GeV}^2$ with $|\mu| \ll {\cal O}(10)$ GeV and $|\mu| \ll {\cal O}(10^7)$ GeV for $|B^2|={\cal O}(10^{-6})~{\rm GeV}^2$ and $|B^2|={\cal O}(1)~{\rm GeV}^2$, respectively.
\item $|\frac{\mu}{\sqrt{2}}v_L| \gg |B^2|$\\
In this case, the mixings are approximately given by
\begin{eqnarray}
s_2 \simeq \frac{R}{\sqrt{1+R^2}},
~~~~
c_2 \simeq \frac{1}{\sqrt{1+R^2}} ,
\label{eq:s2c2}
\end{eqnarray}
where $R=v_{\Phi}/v_L \simeq 10^{-1}$, and the eigenvalues are obtained as
\begin{eqnarray}
m_{h_2}^2 \simeq -\frac{\mu}{\sqrt{2}}v_{\rm ew}\frac{1+R^2}{R},
~~~
m_{h_3}^2 \simeq -\frac{R^2}{1+R^2}\frac{v_{\rm ew}}{v_{\Phi}}B^2 .
\end{eqnarray}
In Eq. (\ref{eq:s2c2}), we ignore contributions from $B^2$.
Given $|\frac{\mu}{\sqrt{2}}v_L + B^2|\simeq|\frac{\mu}{\sqrt{2}}v_L|$ and thus 
${\cal O}(10)~{\rm GeV} < |\mu| < {\cal O}(10^7)~{\rm GeV}$, one obtains ${\cal O}(10^{2})~{\rm GeV} < m_{h_2} < {\cal O}(10^{5})~{\rm GeV}$.
In contrast, the mass of $h_3$ is determined by the symmetry breaking parameter $B^2$; it is roughly given by $m_{h_3}^2 \simeq - B^2 \times 10^8~{\rm GeV}^2$ with $|B^2| \ll {\cal O}(10^{-6})~{\rm GeV}^2$ and $|B^2| \ll {\cal O}(1)~{\rm GeV}^2$ for $|\mu|={\cal O}(10)~{\rm GeV}$ and $|\mu|={\cal O}(10^7)~{\rm GeV}$, respectively.

Here, we would like to stress the point that the modestly large mixing $R \simeq 10^{-1}$ can be gained almost independently of $B^2$.
Thus, the mass of $h_3$ can even be MeV scale while keeping $R \simeq 10^{-1}$.
This feature becomes important when we discuss attenuation of the flux of high energy cosmic neutrinos in Sec. \ref{sec:SnI}.
\end{itemize}

\subsection{Masses of CP-odd neutral scalars}
\label{sec:odd}
In a basis of $(\eta,~\xi,~\sigma_L,~\sigma_{\mu\tau})$, the squared mass matrix of the CP-odd neutral components is written as 
\begin{eqnarray}
M_{\rm odd}^2 =
\bmx{cccc}
-\left( \frac{\mu}{\sqrt{2}} v_L + \sqrt{2}B^2 \right)\frac{v_\Phi}{v_{\rm ew}} & 
\frac{\mu}{\sqrt{2}} v_L + \sqrt{2}B^2 & 
-\frac{\mu}{\sqrt{2}} v_\Phi & 0 
\\
\frac{\mu}{\sqrt{2}} v_L + \sqrt{2}B^2 & 
-\left( \frac{\mu}{\sqrt{2}} v_L + \sqrt{2}B^2 \right)\frac{v_{\rm ew}}{v_\Phi} & 
\frac{\mu}{\sqrt{2}} v_{\rm ew} & 0 \\
-\frac{\mu}{\sqrt{2}} v_\Phi & 
\frac{\mu}{\sqrt{2}} v_{\rm ew} & 
-\frac{\mu}{\sqrt{2}}\frac{v_\Phi}{v_L}v_{\rm ew} & 0 \\
0 & 0 & 0 & 0\\
\emx~.
\end{eqnarray}
Here, $\sigma_{\mu\tau}$ neither mixes with the others nor acquires a non-zero mass; it is the NGB eaten by the $Z_{\mu\tau}$ gauge boson.
Omitting $\sigma_{\mu\tau}$, the diagonalization can approximately be done by
\begin{eqnarray}
\bmx{c} \eta \\ \xi \\ \sigma_L \emx
\rightarrow
\bmx{cccc}
1-\frac{1}{2}r^2 & -r & 0 \\
r & 1-\frac{1}{2}r^2 & 0 \\
0 & 0 & 1 \\
\emx
\bmx{cccc}
1 & 0 & 0 \\
0 & c_2 & s_2 \\
0 &-s_2 & c_2 \\
\emx
\bmx{c} \zeta_1 \\ \zeta_2 \\ \zeta_3 \emx~,
\label{eq:odd-mix}
\end{eqnarray}
where $r=v_\Phi / v_{\rm ew}$.
The mixing $r$ block-diagonalizes the squared mass matrix into a zero eigenvalue and a $2\times 2$ block part
\begin{eqnarray}
m_{\zeta_1}^2 = 0,
~~~
M_{2\times 2} \simeq
\bmx{cc}
-\left( \frac{\mu}{\sqrt{2}} v_L + \sqrt{2}B^2 \right)\frac{v_{\rm ew}}{v_\Phi} & 
\frac{\mu}{\sqrt{2}} v_{\rm ew}
\\
\frac{\mu}{\sqrt{2}} v_{\rm ew} & 
-\frac{\mu}{\sqrt{2}}\frac{v_\Phi}{v_L}v_{\rm ew}
\emx.
\end{eqnarray}
Note that the mass of $\zeta_1^{}$ is exactly vanishing, and it is the NGB eaten by the SM $Z$ gauge boson.
The $2\times 2$ block part $M_{2\times 2}$ has the same form as the $(\phi,s_L)$ block part of Eq. (\ref{eq:Meven}), and it can be diagonalized in the same way as demonstrated in the previous subsection.

\subsection{Invisible decays of the Higgs boson}
\label{sec:inv}
As we have seen in the previous subsections, there are three scalar bosons whose masses can be smaller than half of the Higgs boson mass, namely $h_3^{}$, $h_4^{}$, and $\zeta_3^{}$; these particles cause invisible decays of the Higgs boson $h_1$.
In addition, the Higgs boson can decay into a pair of $Z_{\mu\tau}^{}$ through the mixing between $h_1^{}$ and $h_4^{}$.
We here formulate those interactions as
\begin{eqnarray}
{\cal L}_{\rm inv.}^{} =
&-& 
   G_{h_3}^{}~ h_1^{} h_3^{} h_3^{} 
 - G_{h_4}^{}~ h_1^{} h_4^{} h_4^{} 
 - G_{h_3 h_4}^{}~ h_1^{} h_3^{} h_4^{} 
 - G_{\zeta_3}^{}~ h_1^{} \zeta_3^{} \zeta_3^{} 
\nonumber \\
&-& s_1^{}\frac{M^{2}_{Z_{\mu\tau}}}{v^{}_{\mu\tau}}~ h_1^{} (Z_{\mu\tau}^{})^{\rho}(Z_{\mu\tau}^{})_{\rho}~,
\label{eq:Linv}
\end{eqnarray}
where $\rho$ denotes the Lorentz index and
\begin{eqnarray}
&&G_{h_3^{}}^{} =
  \frac{\mu}{\sqrt{2}}c_1 c_2 s_2
+ \frac{\lambda_3 + \lambda_4}{2}v_{\rm ew}c_1 s_2^2
- \frac{\lambda_7}{2}v_{\mu\tau} s_1 s_2^2
+ \frac{\lambda_9}{2}v_{\rm ew} c_1 c_2^2
- \frac{\lambda_{11}}{2}v_{\mu\tau} s_1 c_2^2,
\\
&&G_{h_4^{}}^{} =
  3\lambda_1^{} v_{\rm ew}^{} c_1^{} s_1^2
- 3\lambda_5^{} v_{\mu\tau}^{} c_1^2 s_1^{}
+ \frac{\lambda_6^{}}{2}\left[
 v_{\rm ew}^{}(c_1^3-2c_1^{}s_1^2)
-v_{\mu\tau}^{}(s_1^3-2c_1^2 s_1^{})
\right],
\\
&&G_{h_3 h_4}^{} = 
  (\lambda_3 + \lambda_4 - \lambda_7)v_\Phi c_1 s_1 s_2
+ (\lambda_9 - \lambda_{11}) v_L c_1 s_1 c_2,
\\
&&G_{\zeta_3^{}}^{} = G_{h_3^{}}^{}.
\end{eqnarray} 
In the above equations, we have neglected $r$ in Eq. (\ref{eq:odd-mix}).
Since $G_{h_3 h_4}$ is suppressed by $v_\Phi$ or $v_L$ in comparison with the others, we will ignore $h_1 \rightarrow h_3 h_4$ decay in the following discussion.

The branching ratio of the Higgs boson invisible decays is constrained by the LHC experiments, and it is defined as 
\begin{eqnarray}
{\rm BR_{inv}}=
\frac{\Gamma_{\rm ex}}{\Gamma_{\rm SM}^{} + \Gamma_{\rm ex}},
\end{eqnarray}
where
\begin{eqnarray}
\Gamma_{\rm ex} =
\sum_{X=h_3, h_4, \zeta_3} \Gamma(h_1^{}\rightarrow XX) 
+ \Gamma(h_1^{}\rightarrow Z_{\mu\tau}^{} Z_{\mu\tau}^{}),
\end{eqnarray}
and $\Gamma_{\rm SM}^{}$ is the SM expectation.
The partial decay widths of $h_1^{}\rightarrow XX$ and
$h_1^{}\rightarrow Z_{\mu\tau}^{} Z_{\mu\tau}^{}$ are calculated to be 
\begin{eqnarray}
&&
\Gamma(h_1^{}\rightarrow XX) = 
\frac{G_X^2}{8\pi m_{h_1}^{}}
\sqrt{1-\left(\frac{2m_{X}^{}}{m_{h_1}^{}}\right)^2},
\\
&&
\Gamma(h_1^{}\rightarrow Z_{\mu\tau}^{} Z_{\mu\tau}^{}) =
\frac{g_{\mu\tau}^2 s_1^2}{8\pi} \frac{M_{Z_{\mu\tau}}^2}{m_{h_1}^{}}
\sqrt{1-\frac{4M_{Z_{\mu\tau}}^2}{m_{h_1}^2}}
\left(
2+\frac{m_{h_1}^4}{4M_{Z_{\mu\tau}}^4}
\left( 1-\frac{2M_{Z_{\mu\tau}}^2}{m_{h_1}^2} \right)^2
\right),
\end{eqnarray}
respectively.
Given the upper limit ${\rm BR_{inv.}}<0.26$ from the ATLAS Collaboration \cite{Aaboud:2019rtt} and $\Gamma_{\rm SM}^{}=4.1$ MeV obtained in Ref. \cite{GamSM}, one finds $\Gamma_{\rm ex} \lesssim 1.44$ MeV.

Let us first estimate constraints from $h_1^{}\rightarrow h_4 h_4$ and $h_1^{}\rightarrow Z_{\mu\tau}^{} Z_{\mu\tau}^{}$.
By substituting Eqs. (\ref{eq:s1c1}) and (\ref{eq:h1h4}), contributions from these decays can be solely expressed in terms of $\lambda_6$ as well as $m_{h_4}^{}$, $g_{\mu\tau}^{}$, and $M_{Z_{\mu\tau}}$.
Then, by requiring that their decay widths should be smaller than $1.44$ MeV, one finds $\lambda_6 < {\cal O}(0.01)$ for $m_{h_4}^{} = {\cal O}(10)$ GeV, $g_{\mu\tau}^{}=5\times 10^{-4}$, and $M_{Z_{\mu\tau}}^{} = 11$ MeV.

Next, depending on $\mu$ and $B^2$, the masses of $h_3$ and $\zeta_3$ can also be smaller than half of the Higgs boson mass, so they cause the Higgs boson invisible decays, too.
We are particularly interested in the case of $|\frac{\mu}{\sqrt{2}}v_L| \gg |B^2|$ as will be discussed in the next section.
In this case, by approximating the mixing with $s_1 \lesssim 10^{-3}$, $c_1 \simeq 1$, $s_2 \simeq 10^{-1}$ and $c_2 \simeq 1$, the constraint $\Gamma_{\rm ex} \lesssim 1.44$ MeV results in $|\mu| < {\cal O}(10)$ GeV, $\lambda_{9}  < {\cal O}(10^{-2})$, $\lambda_{3}, \lambda_{4} < {\cal O}(0.1)$, and $\lambda_7 , \lambda_{11} < {\cal O}(1)$ unless strong cancellation happens in $G_{h_3}$.
Especially, $|\mu| < {\cal O}(10)$ GeV has a strong impact on $\mu$: it fixes the order of $\mu$ to $|\mu|={\cal O}(10)$ GeV in the presence of the lower bound on the charged scalar boson mass, leading to $m_{H^\pm}^{}, m_{h_2}^{}, m_{\zeta_2}^{}={\cal O}(100)$ GeV.

\begin{figure}[th]
\begin{center}
\includegraphics[width=8cm]{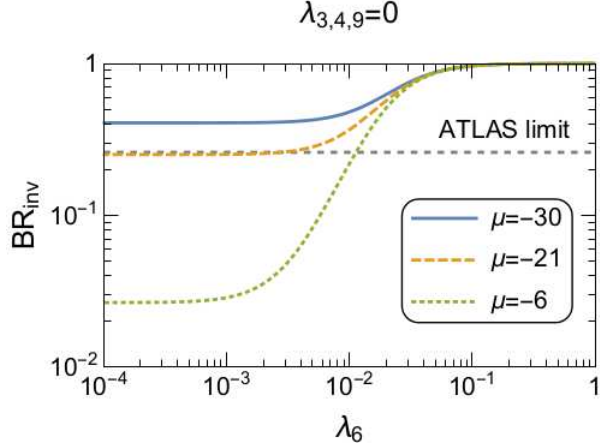}
\end{center}
\vspace{-1cm}
\caption{
The branching ration of the Higgs invisible decay, ${\rm BR_{inv}}$, as a function of $\lambda_6^{}$ for $|\mu| = 30,\ 21$ GeV (a blue-solid, an orange-dashed line) and $|\mu|=6$ GeV (a green-dotted line).
The other parameters are fixed to be $\lambda_{3,4,9}^{}=0$, $m_{h_3}=m_{\zeta_3}=$ 15.6 MeV, $m_{h_4}^{} = $ 28.3 GeV, $g_{\mu\tau}^{}=5\times 10^{-4}$, and $M_{Z_{\mu\tau}}^{} = 11$ MeV.
}
\label{fig:BRinv}
\end{figure}
In Fig. \ref{fig:BRinv}, we plot ${\rm BR_{inv}}$ as a function of $\lambda_6^{}$ for $|\mu| = 30,\ 21$ GeV (a blue-solid, an orange-dashed line) and $|\mu|=6$ GeV (a green-dotted line); the other parameters are fixed to be $\lambda_{3,4,9}^{}=0$, $m_{h_3} = m_{\zeta_3}=$ 15.6 MeV, $m_{h_4}^{} = $ 28.3 GeV, $g_{\mu\tau}^{}=5\times 10^{-4}$, and $M_{Z_{\mu\tau}}^{} = 11$ MeV.
From the figure, one can see that $|\mu| \lesssim 21$ GeV as well as $\lambda_6 \lesssim 0.01$.

\section{Neutrino secret interactions}
\label{sec:SnI}
Lastly, we briefly comment on the so-called neutrino secret interactions.
Soon after the discovery of high energy cosmic neutrino events in the energy range between ${\cal O}(100)$ TeV and ${\cal O}(1)$ PeV by the IceCube Collaboration \cite{Aartsen:2013bka,Aartsen:2013jdh}, several authors pointed out that if there exist new light bosons having interactions with neutrinos, they could cause significant attenuation in the flux of high energy cosmic neutrinos by mediating resonant scattering with cosmic neutrino background  
\cite{Ioka:2014kca,Ng:2014pca,Ibe:2014pja,Araki:2014ona,Kamada:2015era,DiFranzo:2015qea,Araki:2015mya}.
Interestingly, one can indeed find a possible dip in the energy spectrum of high energy cosmic neutrinos around $400$ TeV $-$ $1$ PeV in the IceCube data, see Ref. \cite{Kopper:2017zzm} for the recent analysis.
In our models, we have three candidates for such light bosons, that is, the CP-even neutral scalar $h_3^{}$ and the CP-odd one $\zeta_3^{}$, as well as $\Zp$.
The detailed analysis for $\Zp$ was already done in Refs. \cite{Araki:2014ona,Araki:2015mya}.
Thus, in this section, we focus only on $h_3^{}$ and $\zeta_3^{}$ and check whether they can serve as the mediator of the resonant scattering, just as $\Zp$ does.

Suppose the light boson is scalar and has Yukawa interactions with active neutrinos, the interaction can be formulated as 
\begin{eqnarray}
g_\nu^{} \bar{\nu} \nu^c \chi + h.c. ~.
\label{eq:geff}
\end{eqnarray}
According to Ref. \cite{Araki:2015mya}, in order for the scalar $\chi$ to significantly attenuate the cosmic neutrino flux in the PeV region, its mass and the coupling constant should satisfy
\begin{eqnarray}
M_\chi^{} = {\cal O}(1-10)~{\rm MeV} ,~~
g_\nu^{} > {\cal O}(10^{-4}) ~.
\end{eqnarray}
As shown in Sec. \ref{sec:scalar}, both $h_3^{}$ and $\zeta_3^{}$ can have an MeV scale mass by suitably tuning the $U(1)_L^{}$ breaking parameters $B^2$.
As for the Yukawa coupling $g_\nu^{}$, in our models, it arises from the third term in Eq. (\ref{eq:Lag}), through the mixing in Eqs. (\ref{eq:AS-mix}) and (\ref{eq:s2c2}).
In the case of $h_3^{}$, it arises as 
\begin{eqnarray}
y_N^{} \overline{\nu_L^{}} (N_L^{})^{c} \phi
\rightarrow
-y_N^{} 
\left[ 
\left(1-\frac{1}{2}\eta\eta^\dag \right) \eta^T_{} R
\right] 
\overline{\nu} \nu^c h_3^{} 
\equiv g_\nu^{} \overline{\nu} \nu^c h_3^{} ,
\end{eqnarray}
while for $\zeta_3^{}$, it is described as
\begin{eqnarray}
i y_N^{} \overline{\nu_L^{}} (N_L^{})^{c} \xi
\rightarrow
-iy_N^{} 
\left[ 
\left(1-\frac{1}{2}\eta\eta^\dag \right) \eta^T_{} 
R\left(1-\frac{1}{2}r^2\right) 
\right] 
\overline{\nu} \nu^c \zeta_3^{} 
\equiv g_\nu^\prime \overline{\nu} \nu^c \zeta_3^{} .
\end{eqnarray}
Given our parameter setting: $\eta={\cal O}(10^{-2})$, $R=v_\Phi/v_L^{}={\cal O}(10^{-1})$, and $r=v_\Phi/v_{\rm ew}^{}={\cal O}(10^{-10})$, one obtains $g_\nu^{(\prime)}={\cal O}(10^{-3})$, which is presumably large enough to attenuate the cosmic neutrino flux.
A more detailed study will be done elsewhere.

\section{Summary}
\label{sec:sum}
In summary, we propose models for an MeV scale $U(1)_{\lmlt}^{}$ gauge boson, $\Zp$, which is anticipated to exist to resolve the $g_\mu^{} -2$ problem as well as the possible dip in the energy spectrum of high energy cosmic neutrinos.
We introduce extra scalars to spontaneously break $U(1)_{\lmlt}^{}$ and to generate an MeV scale mass to $\Zp$.
Tiny neutrino masses and mixing are obtained by simultaneously invoking the linear and the inverse seesaw mechanism.
Depending on the $U(1)_{\lmlt}^{}$ charge assignment, the active neutrino mass matrix enjoys two types of one-zero textures, which endows the models with predictive power.
Both of the textures prefer inverted neutrino mass ordering and would be tested in next generation experiments of neutrinoless double beta decay.
Furthermore, we find that two of the extra scalars can acquire MeV scale masses while having interactions with active neutrinos.
We briefly confirm that they can serve as the mediator of resonant scattering between high energy cosmic neutrinos and cosmic neutrino background, and that they could help us understand the existence of the unexpected dip in the IceCube data.

Finally, we comment on kinetic mixing between $U(1)_{\lmlt}$ and the SM electromagnetic $U(1)_{\rm em}$ gauge symmetry.
In our framework, we do not introduce the kinetic mixing term at the tree level, but it appears at the one loop level.
In Refs. \cite{Araki:2017wyg,Araki:2015mya}, we studied implications of the loop-induced kinetic mixing for solar neutrino measurements and the detectability of $\Zp$ at $e^+ e^-$ colliders.
In contrast to the previous work, we here introduce extra fermions and scalars charged under $U(1)_{\lmlt}$.
Nevertheless, there are no mass eigenstates that are charged under both $U(1)_{\lmlt}$ and $U(1)_{\rm em}$, except for the SM mu and tau leptons.
Hence, we expect that the same kinetic mixing as those in Refs. \cite{Araki:2017wyg,Araki:2015mya} will also be obtained for the framework proposed in this paper.

\begin{acknowledgments}
This work was supported by JSPS KAKENHI Grants No.~JP18H01210 (T.A., J.S., T.S.), 
No.~JP19J13812 (K.A.), 
No.~JP18K03651 (T.S.), 
and MEXT KAKENHI Grant No.~JP18H05543 (J.S., T.S.).
\end{acknowledgments} 

\bibliography{./Lmu-Ltau_Seesaws}
\bibliographystyle{apsrev4-1}

\end{document}